
\magnification 1200
\headline{\hfil\tenrm\folio\hfil}
\footline={\hfil}
\headline{\ifnum\pageno=1\nopagenumbers
\else\hss\number\pageno\fi}
\footline={\hfil}
\overfullrule=0pt
\baselineskip=20pt
 \hfuzz=5pt \null

\def\R{ {\rm R \kern -.31cm I \kern .15cm}}
\vskip 1 cm
\centerline{{\bf STRUCTURE FUNCTIONS AND LOW X PHYSICS}}
\par
\bigskip
\centerline {by}
\medskip
\centerline{{\bf A. Capella, A. Kaidalov}\footnote{$^+$}{Permanent address :
ITEP, Moscow (Russia).},
{\bf C. Merino and J. Tran Thanh Van}} \centerline{Laboratoire de Physique
Th\'eorique et Hautes Energies\footnote{*}{Laboratoire
associ\'e au Centre National de la Recherche
Scientifique - URA 63.}} \centerline{B\^atiment 211, Universit\'e
de Paris-Sud, 91405 Orsay cedex, France \footnote{}{\noindent To receive copies
of the figures,
please contact the authors at Merinoatqcd.th.u-psud.fr}}  \vskip1cm

\noindent \underbar{\bf Abstract} \par
In the framework of conventional Regge theory we present a common description
of total
photon-proton cross-section and proton structure functions in the region $0
\leq Q^2 \leq 5 \
{\rm GeV}^2$. Using it as an initial condition in the perturbative QCD
evolution equation, we
describe structure functions at any $Q^2$ and $x$ - including the small $x$
region explored
at HERA.

\vskip2cm
 \noindent{\obeylines LPTHE Orsay 94-34
\noindent April 1994}
\vfill\eject
\baselineskip=20pt

Perturbative QCD predicts the $Q^2$-dependence of structure functions of
hadrons and nuclei when
the initial conditions (i.e. the momentum distribution functions of quarks and
gluons at $Q^2 =
Q_0^2$) are given. The value of $Q_0^2$ has to be large enough for the
perturbative expansion
(which is known in two-loops) to be valid. The determination of the initial
condition is a problem
of soft (non-perturbative) physics. This problem is especially important in the
region of very
small $x$ which is now experimentally accessible at HERA. Indeed, the study of
the region $x << 1$
(Regge limit) and large $Q^2$, should help to clarify the interplay between
perturbative and
non-perturbative physics. \par

In this letter, we use the present knowledge of soft diffractive processes to
formulate the initial
condition for perturbative QCD evolution. Our aim is to obtain a simple,
theoretically motivated
description of both the total $\gamma p$ cross-section and the structure
function in a region of low
and moderate values of $Q^2$ ($0 \leq Q^2 \leq 5 \ {\rm GeV}^2$), covering the
full $x$-region.
Using this initial condition in the perturbative QCD evolution equation we
obtain the
structure functions at larger values of $Q^2$. Our results are in agreement
with experiment in the
whole $Q^2$ and $x$ region. \par

\vskip 5 mm
\noindent \underbar{\bf Small x behaviour of the structure functions} \quad In
Regge theory the
high-energy behaviour of hadron-hadron and photon-hadron total cross-sections
is determined by the
Pomeron intercept $\alpha_P = 1 + \Delta$, and is given by
$\sigma^{tot}_{\gamma (h) p}(\nu) \sim
\nu^{\Delta}$. This behaviour is also valid for a virtual photon for $x =
Q^2/2m \nu << 1$, leading
to the well known behaviour, $F_2(x, Q^2) \sim x^{- \Delta}$, of the structure
functions at fixed
$Q^2$ and $x \to 0$. \par

In many analysis of the structure functions either $\Delta = 0$ or $\Delta =
0.5$ have been
used$^{[1]}$. The first value corresponds to a constant cross-section and
sometimes it is wrongly
as\-so\-cia\-ted with the intercept of the soft Pomeron. The second value
corresponds to an estimate
of the hard (Lipatov) Pomeron intercept. Both choices are in disagreement with
our knowledge of the
Pomeron intercept inferred from a study of the energy dependence of hadronic
cross-sections. This
behaviour is universal and given by $\nu^{\Delta_0}$ with $\Delta_0 \approx
0.08^{[2]}$. However,
due to the existence of absorptive corrections (rescattering), this is not the
true Pomeron
intercept (sometimes referred to as ``bare'' Pomeron), but rather an effective
one. The ``bare''
Pomeron intercept is in fact substantially larger than $1 + \Delta_0$. In the
generalized eikonal
model (with diffractive intermediate states taken into account) one gets
$\Delta_1 = 0.12 \div
0.14^{[3, 4]}$. In a theory containing a more complete set of absorptive
graphs, one obtains
an even larger value $\Delta_2 = 0.2 \div 0.25^{[5]}$, leading to the ``bare''
Pomeron intercept $1 +
\Delta_2$. These considerations are very important if one attempts to cover a
broad region of $Q^2$
which includes also the real photon ($Q^2 = 0$). Indeed the relative
contribution of the most
important absorptive corrections decreases quite rapidly when $Q^2$ increases
(like $Q^{-2}$ for the
eikonal ones), so that as $Q^2 \to \infty$ we see a Pomeron intercept which is
close to the ``bare''
one $1 + \Delta_2$ - much larger than the effective intercept $1 + \Delta_0$
seen at $Q^2 = 0$. These
considerations have prompted us to use a low $x$ behavior of the structure
function of the form

$$F_2(x, Q^2) \sim x^{-\Delta(Q^2)} \quad ; \quad \Delta(Q^2) = \Delta_0 \left
( 1 + {2 Q^2 \over
Q^2 + d} \right ) \eqno(1)$$

\noindent where $\Delta_0$ and $d$ are free parameters. In the fit described
below one obtains
$\Delta(Q^2 = 0) = \Delta_0 \sim 0.08$, so that $\Delta(Q^2 \to \infty) \sim
0.24$, corresponding
to the ``bare'' Pomeron intercept $1 + \Delta_2$, as discussed above. \par

\vskip 5 mm
\noindent \underbar{\bf Structure functions at low $Q^2$} \quad In order to
obtain the initial
condition for perturbative QCD evolution we have to know the structure
functions at $Q^2 = Q_0^2$
and all values of $x$. Furthermore it is necessary to know separately the
contributions of valence
and sea quarks and gluons. According to the large-$N$ expansion, we assume that
the Pomeron
determines the small-$x$ behaviour of sea quarks and gluons (eq. (1)), while
secondary reggeons
($\rho , \omega , f, A_2$), with intercept $\alpha_R \sim 0.4 \div 0.6$,
determine that of valence
quarks. \par

In order to include real photons ($Q^2 = 0$), we have to use the relation

$$\sigma_{\gamma p}^{tot}(\nu) = \left [ {4 \pi^2 \alpha_{EM} \over Q^2} F_2(x,
Q^2) \right ]_{Q^2=0}
\ \ \ , \eqno(2)$$

\noindent which implies that $F_2(x, Q^2) \sim Q^2$ at $Q^2 \to 0$. With these
considerations in
mind we propose the following parametrization of the structure functions in the
region of small
and moderate $Q^2 (0 \leq Q^2 \leq 5 \ {\rm GeV}^2$)~:

$$F_2(x, Q^2) = A \ x^{-\Delta(Q^2)} (1 - x)^{n(Q^2)+4} \left ( {Q^2 \over Q^2
+ a} \right
)^{1+\Delta(Q^2)}$$
$$+ B \ x^{1 - \alpha_R} (1 - x)^{n(Q^2)} \left ( {Q^2 \over Q^2 + b} \right
)^{\alpha_R} \ \ \ .
\eqno(3)$$

\noindent The first term in (3) corresponds to the Pomeron contribution with an
$x \to 0$
behaviour given by eq. (1). The second term corresponds to the secondary
reggeon contribution
with an $x \to 0$ behaviour determined by the secondary reggeon intercept
$\alpha_R$. The behaviour
at $x \to 1$ is given by the second factor, with $n(Q^2)$ parametrized as

$$n(Q^2) = {3 \over 2} \left ( 1 + {Q^2 \over Q^2 + c} \right ) \ \ \ ,
\eqno(4)$$

\noindent in such a way that, at $Q^2 = 0$, we have the same power $(1 -
x)^{1.5}$ as in the Dual
parton model$^{[6]}$ (also controlled by Regge intercepts~: $n(0) = \alpha_R(0)
- 2 \alpha_N(0)
\approx {3 \over 2}$), and at $Q^2 \to \infty$ we have $(1 - x)^3$ as given by
dimensional counting
rules. The last factor in the two terms of eq. (3) is
required by eq. (2) in order to connect with real photons. (A similar factor
has been
previously introduced in the same physical context ; see [7], [8] and
references therein).
Using eq. (2) one has~:

$$\sigma_{\gamma p}^{tot}(\nu) = 4 \pi^2 \alpha_{EM} \left ( A \
a^{-1-\Delta_0}(2
m \nu)^{\Delta_0} + B \ b^{-\alpha_R}(2 m \nu)^{\alpha_R - 1} \right ) \ \ \ .
\eqno(5)$$

In this way we have a parametrization of both structure functions and $\gamma
p$ total
cross-section with 8 free parameters. Four of these parameters appear in (5) so
that the structure
function of a proton contains only 4 extra parameters. Out of these only three
are free since we fix
the parameter $B$ using the normalization condition for valence quarks. In
order to do so we have
first to separate in the second term of (3) the contribution of the $u$ and $d$
valence quarks.
This is done by replacing the term $B(1 - x)^{n(Q^2)}$ in eq. (3) as follows~:

$$B(1 - x)^{n(Q^2)} \to B_u(1 - x)^{n(Q^2)} + B_d(1 - x)^{n(Q^2)+1} \ \ \ .
\eqno(6)$$

The condition of having in the proton two valence $u$-quarks (charge 2/3) and
one valence $d$-quark
(charge $-1/3$), determines the values of $B_u$ and $B_d$ (see below). \par

The seven remaining parameters were determined from a joint fit of the
$\sigma_{\gamma
p}^{tot}$ data and the new NMC data$^{[9]}$ on the proton structure function
$F_2(x,
Q^2)$ in the region $1 \ {\rm GeV}^2 \leq Q^2 \leq 5 \ {\rm GeV}^2$. The
description of the data is
quite good (the $\chi^2$ is practically the same as the one of the
parametrization used in ref. [9]
with 15 free parameters, and much better than the one of the parametrization of
ref. [8] with
20 free parameters). Our results are shown in Figs. 1 and 2 and the values of
the parameters
(which are strongly correlated) are given in the caption of Fig. 2. A
comparison of our predictions
for $F_2(x, Q^2)$ with the SLAC data$^{[10]}$ (not included in the fit) is
shown in Fig. 3. The
agreement is quite good. \par

As mentioned above our parametrization gives the separate contributions of
valence and sea quarks.
Thus, we can predict the structure function $F_2(x, Q^2)$ of a neutron
(deuteron). In Fig. 4 we
compare our predictions for the structure function of a deuteron with
experimental data$^{[9, 10]}$.
The agreement is very good. Likewise, structure functions for $\nu (\bar{\nu}
)N$-scattering are
also well described$^{[11]}$. \par

In order to determine the distribution of gluons in a nucleon we assume that
the only difference
between distributions of sea-quarks and gluons is in the $x \to 1$ behaviour.
Using the arguments
of ref. [12] we write it in the form

$$x \ g(x, Q^2) = G x \bar{q}(x, Q^2)/(1 - x) \ \ \ , \eqno(7)$$

\noindent where $x \bar{q}(x, Q^2)$ is proportional to the first term of eq.
(3). The constant
$G$ is determined from the energy-momentum conservation sum rule. Comparison of
our
predictions with the available data is shown in Fig. 5.

\vskip 5 mm
\noindent \underbar{\bf Structure functions at large $Q^2$} \quad We are now
ready to introduce the
QCD-evolution in our partonic distributions and thus to determine structure
functions at larger
values of $Q^2$. We use the evolution equation in two loops in the
$\overline{MS}$ scheme with
$\Lambda = 200 \ {\rm MeV}$. \par

For the initial point $Q_0^2$ we can use any value in the region $1 \ {\rm
GeV}^2 \leq Q^2 \leq 5 \
{\rm GeV}^2$. However, since our initial condition describes the data in the
whole region $Q^2 \leq 5
\ {\rm GeV}^2$, we know not only the function $F_2(x, Q_0^2)$ but also its
derivative $\left .
{dF_2(x, Q^2) \over d \ \ell n Q^2} \right |_{Q^2=Q_0^2}$, which in general
will not coincide with
that obtained from the QCD evolution equation. We checked, however, that for
$Q_0^2 \approx 2 \ {\rm
GeV}^2$, these two derivatives are very close to each other in a broad region
of $x$. So we have
chosen as an initial condition $Q_0^2 = 2 \ {\rm GeV}^2$ in order to obtain a
smooth behaviour in
$Q^2$. Actually, it is possible to get an exact equality of derivatives if, for
$Q^2 \geq Q_0^2$,
we add to the QCD-evolution a term decreasing as a power of $Q^2$ (higher
twist)~:

$$F_2(x, Q^2) = F_2^{pert}(x, Q^2) \left ( 1 + {f(x) \over Q^2} \right ) \qquad
Q^2 \geq Q_0^2 \
\ \ . \eqno(8)$$

\noindent The function $f(x)$ is determined by requiring the equality of the
$Q^2$-derivatives
of our initial condition (3) and of eq. (8) at the point $Q_0^2$. The results
obtained using (8) are
shown in Figs. 2-4  as full curves. The agreement with the data is quite good.
\par

Although our fit of the NMC data has been restricted to the region $1 \ {\rm
GeV}^2 \leq Q^2 \leq 5 \
{\rm GeV}^2$, it is interesting to mention that a good fit can also be obtained
up to $Q^2
\sim 10 \ {\rm GeV}^2$. This produces only small changes in our parameters and
allows to start the
perturbative QCD evolution at larger values of $Q^2$. Details of such a
parametrization will be
given elsewhere. \par

\vskip 5 mm
\noindent \underbar{\bf The low x HERA region} \quad Our predictions for the
small $x$-region
measured at HE\-RA$^{[13,14]}$ are shown in Figs. 6, 7. It follows from these
figures that the
initial condition, eq. (3), based on our knowledge of the Pomeron properties
inferred from soft
processes, together with conventional QCD evolution, can explain the increase
at very small $x$
observed at HERA$^{[13, 14]}$. \par

This is obtained with a Pomeron intercept $1 + \Delta(Q^2)$ with $\Delta (Q^2
\to \infty ) = 0.2
\div 0.25$. This is about half way between the values $\Delta = 0$ and $\Delta
= 0.5$ used in
the literature at large $Q^2$. Since this value of the Pomeron intercept is
consistent with the
one obtained in the study of high energy hadronic interactions, we consider
that our results
are in favor of the idea$^{[6, 15-17]}$ that there is only one Pomeron (rather
than a soft and a
hard ones). For discussion on this point see for example ref. [18]. \par

In our approach the low-$x$ increase of the structure functions observed at
HERA is largely due
to the initial condition. This approach is orthogonal to the one in ref. [19]
where the low-$x$ rise
at HERA is entirely due to the perturbative QCD evolution (started at $Q^2 =
Q_0^2 \sim 0.3 \ {\rm
GeV}^2$). An intermediate situation occurs in the approaches of ref. [20].
Forthcoming data from the
E665 collaboration at Fermilab in the unexplored region of low $x$ and low
$Q^2$ should
provide a clear cut distinction among these approaches.

\vfill\eject
\centerline{\underbar{\bf References}} \bigskip
\item {1.} B. Badelek, K. Charchula, M. Krawczyk and J. Kwiecinski, Rev. Mod.
Phys. \underbar{64}
(1992) 927. \item{2.} A. Donnachie, P. V. Landshoff, Phys. Lett.
\underbar{B296} (1992) 227. \par
\item {3.} A. Capella, J. Kaplan and J. Tran Thanh Van, Nucl. Phys.
\underbar{B97} (1975) 493.\par
\item{4.} K. A. Ter-Martirosyan, Sov. J. Nucl. Phys. \underbar{44} (1986)
817.\par
\item{} A. B. Kaidalov, K. A. Ter-Martirosyan and Yu. M. Shabelski, Sov. J.
Nucl. Phys. \underbar{43}
(1986) 822. \par  \item {5.} A. B. Kaidalov, L. A. Ponomarev and K. A.
Ter-Martirosyan, Sov. J. Nucl.
Phys. \underbar{44} (1986) 468.\par \item {6.} A. Capella, U. Sukhatme, C-I Tan
and J. Tran Thanh
Van, Phys. Reports \underbar{236} (1994) 225. \par \item {7.} A. Donnachie, P.
V. Landshoff, Z. Phys.
\underbar{C61} (1994) 139. \par \item {8.} H. Abramowicz, E. M. Levin, A. Levy
and U. Maor, Phys.
Letters \underbar{B269} (1991) 465. \par  \item {9.} P. Amaudruz et al. (NMC),
Phys. Lett.
\underbar{B295} (1992) 159. \par  \item {10.} L. W. Whitlow et al., Phys. Lett.
B \underbar{282}
(1992) 475. \par \item{11.} A. Capella, A. Kaidalov, C. Merino and J. Tran
Thanh Van, Proc. of XXIV
Rencontre de Moriond, M\'eribel, France (1994). To be published. \par
\item {12.} F. Martin, Phys. Rev. \underbar{D19} (1979) 1382. See also F.
Indur\'ain and C. L\'opez,
Nucl. Phys. \underbar{B171} (1980) 231.   \par  \item {13.} I. Abt et al. (H1
Collaboration), Nucl.
Phys. \underbar{B407} (1993) 515. \par \item {14.} M. Derrick et al. (ZEUS
Collaboration), Phys.
Lett. \underbar{B316} (1993) 412. \par \item{15.} A. Kaidalov in Proc. XIXth
Inter. Symposium on
Multiparticle Dynamics, Arles, France (1988), Ed. Fronti\`eres and World
Scientific. \par
\item{16.} P. V. Landshoff in Proc. of 3rd Int. Conference on Elastic and
Diffractive Scattering,
Nucl. Phys. \underbar{B12} (1990) 397. \par
\item{17.} E. Levin and C-I Tan, Brown HET-889 (1992). \par
\item{18.} J. Bjorken in Proc. of 4th Int. Conference on Elastic and
Diffractive Scattering, Italy
(1991). \par
\item{19.} M. Gl\"uck, E. Reya, A. Vogt, Z. Phys. \underbar{C53} (1992) 127.
\par
\item{20.} B. Badelek and J. Kwiecinski, Phys. Lett. \underbar{B295} (1992)
263-268. \par \item{}
M. Bertini et al., LYCEN 9366 (1993).  \par

\vfill \supereject
\centerline{\underbar{\bf Figures Captions}} \bigskip
{\parindent = 1 truecm
\item{\bf Fig. 1 } The data for total $\gamma p$ cross-section versus
$\sqrt{s}$, including the
high-energy HERA ones, are compared with the result of a joint fit of
$\sigma_{\gamma p}$ and NMC
proton structure functions described in the main text. \par
\item{\bf Fig. 2} The proton structure function $F_2(x, Q^2)$ obtained by the
NMC
collaboration$^{[9]}$ at 90 GeV (full circles) and 280 GeV (open circles). The
dashed line is
obtained from eqs. (3) and (6) with the following values of the parameters (all
dimensional
parameters are in GeV$^{2}$)~: $A = 0.1502$, $a = 0.2631$, $\Delta_0 =
0.07684$, $d = 1.1170$, $b
= 0.6452$, $\alpha_R = 0.4150$, $c = 3.5489$. (The values of the parameters
$B_u$ and $B_d$ in
(6), determined from the normalization condition for valence quarks (at $Q^2 =
2 \ {\rm GeV}^2$),
are~ : $B_u = 1.2064$ and $B_d = 0.1798$. The full line is obtained using QCD
evolution (see eq.
(8)). \par
\item{\bf Fig. 3} Same as in Fig. 2 except that the data are now from
SLAC$^{[10]}$. \par
\item{\bf Fig. 4} Same as in Figs. 2 and 3 for the deuteron structure
functions. Here $B_u = 0.7540$
and $B_d = 0.4495$. \par
\item{\bf Fig. 5} The gluon structure function at $Q^2 = 9 \ {\rm GeV}^2$. The
theoretical curve is
obtained from eq. (7). \par
\item{\bf Fig. 6} The $x$-dependence of the proton structure function at three
different values of
$Q^2$. The lower (upper) curve for $Q^2 = 15 \ {\rm GeV}^2$ is obtained using
eq. (8), with
(without) the higher twist term. \par
\item{\bf Fig. 7} The $Q^2$-dependence of the proton structure function for
different values of $x$,
compared with HERA data. \par
}

\bye